\documentstyle[twocolumn,aps,epsfig]{revtex}
\begin{document}
\title{Interplay of linear and modified nonlinear impurities in the 
formation of stationary localized states}

\author{Bikash C. Gupta and Sang Bub Lee}
\address{Department of Physics, Kyungpook National University,
Taegu 702--701, Korea}

\maketitle
\begin{abstract}
Formation of stationary localized states in one-dimensional chain 
due to a linear impurity and a modified nonlinear 
impurity is studied.
Furthermore, a one-dimensional chain with linear and modified nonlinear 
site energies at the alternate sites is studied and
rich phase diagrams of SL states are obtained for all the systems we 
considered.
The results are compared with those of the linear and nonlinear systems.   
\end{abstract}
\pacs{PACS numbers : 71.55.-i, 72.10.Fk}

\section{introduction}

It is well known that the transport properties of a system is directly
related to the formation of stationary localized (SL) states in the system.
Localized states appear due to the presence of impurity or disorder 
(which breaks the translational symmetry) in the system \cite{01}.
There has been detailed study on the formation of SL states in various
systems in the presence of linear impurities. One the other hand the
effect on the formation of localized states due to the presence of 
nonlinear impurities has been studied recently by the use of the 
discrete nonlinear Schr\"odinger equation given as 
\cite{02,03,04,05,06,07,08,09,10,11,12,13,14,15,16}
\begin{equation}
i \frac{d C_{n}}{dt} =  \epsilon_n C_n  + V (C_{n+1} + C_{n-1}) 
+ \chi_{n}  | C_{n} |^\sigma C_{n} ,
\end{equation}
where $\sigma$ is equal to 2 or arbitrary constant, $C_n$ is the 
probability amplitude of the particle to be at the site $n$, $\epsilon_n$
and $\chi_n$ are the static site energy and the nonlinear strength at
site $n$ respectively, and $V$ is the nearest neighbor hopping element.
The nonlinear term $|C_n|^\sigma C_n$ arises due to interaction of the
exciton with the lattice vibration \cite{17,18}. If the lattice oscillators 
are purely harmonic in nature, then $\sigma = 2$. The Eq.~(1) has been
derived with the assumption that the lattice oscillators are much
faster in motion compared to that of the exciton and the lattice
oscillators are independent of others. Equation ~(1) has been used
to study the formation of SL states due to a single nonlinear impurity
and that due to a dimeric nonlinear impurity in one dimensional chain
as well as in a Cayley tree. A perfect nonlinear chain with nonlinear 
impurity have also been considered to study the SL states and it has 
been shown that SL states appear even though the translational symmetry
of the system is preserved \cite{10,11,12,13,14,15,16}. Very recently 
the effect of the presence of both the linear and nonlinear impurities 
in one dimensional chain and in the Cayley tree has been studied and 
rich phase diagrams of SL states have been obtained \cite{19}. 
The Eq. ~(1) may be modified if the lattice
oscillators are assumed to be coupled with their nearest neighbors.
Thus, under the assumption that the oscillators are coupled to its 
nearest neighbors and are much faster in motion compared to the
motion of the exciton, the discrete nonlinear Schr\"odinger equation
is modified and takes the form:
\begin{eqnarray}
i\frac{dC_n}{dt} = \epsilon_n C_n + V (C_{n+1} + C_{n-1}) \nonumber \\
+ \chi_n (|C_{n+1}|^2 + |C_{n-1}|^2 + 2|C_n|^2) C_n,
\end{eqnarray}
where $C_n$, $\epsilon_n$, $V$ and $\chi_n$ carries same meaning as 
for the Eq.~(1). The modified discrete nonlinear Schr\"odinger equation
(Eq. ~(2)) has been derived from the coupled exciton-oscillator system
by Kopidakis et al. \cite{20}. We notice that the Eq. (2) contains 
more nonlinear terms compared to Eq.~(1) and, in addition, Eq.~(2) is 
also relevant in condensed matter physics because the nearest neighbor
coupling among the lattice oscillators is taken into account in 
deriving the equation. One interesting property of Eq.~(2) is the
self-trapping property and this aspect has been studied by Kolosakas
et al. \cite{21}. Another interesting property of this equation is
that it can produce SL states. In fact the Eq.~(2) has been exploited 
to study the formation
of SL states in one dimensional system in the presence of a single
nonlinear impurity, and that in the presence of a dimeric nonlinear
impurity \cite{22}. The Eq. ~(2) has also been used to find the SL 
states in a perfectly nonlinear chain \cite{22}. However the system 
with both the linear and the modified 
nonlinear impurities has not been considered yet to study the formation
of SL states. Thus we are interested to observe the mixed role of linear
and modified nonlinear impurities on the formation of SL states in
one dimensional system.

The organization of the paper is as follows. The one dimensional chain
with a linear impurity and a modified nonlinear impurity is considered
in Sec. I. In Sec. II we consider the one dimensional chain with linear
and modified nonlinear impurities at alternate sites of the lattice. 
Finally In Sec. III we summarize our findings.

\section{one dimensional chain with a linear impurity and a modified 
nonlinear impurity}

Consider a one-dimensional chain consisting of a linear impurity of strength
$\epsilon$ placed at the zeroth site and a nonlinear impurity of strength 
$\chi$ at the first site. The relevant Hamiltonian for such a system can be 
written as
\begin{eqnarray}
H = \frac{1}{2} \sum_{n= - \infty}^{\infty} 
(C_{n} C_{n+1}^{\star} + C_{n}^{\star} C_{n+1}) 
 + \frac{\epsilon}{2} | C_{0} |^{2} \nonumber \\
+ \frac{\chi}{4} \left(|C_2|^2 + |C_0|^2 + 2|C_1|^2 \right) |C_1|^2 ,
\end{eqnarray}
where the nearest-neighbor hopping matrix element is assumed to be unity.
As a possible solution for the stationary localized states, we consider
\begin{equation}
C_{n} = \phi_{n} e^{-iEt} ,
\end{equation}
where $E$ is the stationary state energy. It is also well known that impurity 
states in one-dimensional system are exponentially localized.  Therefore, the 
presence of two impurities at the two consecutive sites (site 0 and site 1) 
in the chain, suggests us to consider the following forms of $\phi_{n}$:
\begin{eqnarray}
& &\phi_{n} =  \phi_{1} \left[ {\rm sgn}(E) \eta \right]^{n-1} ~~~~~~~~ 
n \ge 1 \nonumber \\
& &\phi_{-|n|}  =  \phi_{0} \left[ {\rm sgn}(E) \eta \right]^{|n|} 
~~~~~~~ n \le 0 . 
\end{eqnarray}
The form of $\phi_{n}$ given above are exact and may be derived from the
Green's function analysis\cite{07}.
Here $\eta$ lies between 0 and 1 and is given as
\begin{equation}
\eta = \frac{1}{2} \left[ |E| - \sqrt{E^{2} - 4} \right] .
\end{equation}
Since we are dealing with the localized states, $\phi_{0}$ and $\phi_{1}$
can be assumed to be real without forsaking the mathematical rigor.
Three possible cases may be considered, i.e.,
$\phi_{1} = \phi_{0}$, $\phi_{1} = - \phi_{0}$, and $\phi_{1} \ne \phi_{0}$,
the solutions for which are called, respectively, symmetric,
antisymmetric, and asymmetric solutions.
To take into account of all three possibilities,
we introduce a variable $\beta = \phi_{0} / \phi_{1}$.
Consequently, $\beta = {1}$ corresponds to the symmetric solutions,
$\beta = -1$ to the antisymmetric solutions, and $\beta \ne \pm 1$
to the asymmetric solutions.
Substituting the ansatz of Eq.~(5) and the definition of $\beta$
to the normalization condition $\sum | C_{n} |^{2} = 1$, we obtain
\begin{equation}
| \phi_{0} |^{2} = \frac{1- \eta^{2}}{1 + \beta^{2}}.
\end{equation}
Using Eqs.~(4), (5) and (7) in Eq.~(3),
we obtain the effective Hamiltonian for the reduced dynamical system:
\begin{eqnarray}
H_{\rm eff} = {\rm sgn}(E) \eta + \frac{(1 - \eta^2 )}{(1 + \beta^2)}
\left( \beta + \frac{\epsilon}{2}\right) \nonumber \\
+ \frac{\chi}{4} 
\frac{(1 - \eta^2)^2}{(1 + \beta^2)^2} (\beta^{4}\eta^2 + 2\beta^4 + \beta^2).
\end{eqnarray}

We first consider the case of $\beta = \pm 1$.
The effective Hamiltonian for this case becomes
\begin{eqnarray}
H_{\rm eff}^{\pm} = {\rm sgn}(E) \eta \pm \frac{(1 - \eta^{2})}{2} 
+ \frac{\epsilon}{4} (1 - \eta^2) \nonumber \\
+ \frac{\chi}{16} (3+\eta^2)(1 - \eta^{2} )^{2} ,
\end{eqnarray}
where ``$+$'' refers to the symmetric states and ``$-$'' corresponds to the
anti-symmetric states.
For a particular system, $\epsilon$ and $\chi$ are constants,
while $\eta$ is a variable which determines the energy of the localized state.
The value of $\eta$ should be determined self-consistently and,
for this reason, we treat $\eta$ as the dynamic variable in the
effective Hamiltonian of the reduced dynamical system.
The fixed-point solutions corresponding to the localized states
can be found, from the variational principle, by the condition
$\partial H_{\rm eff}^{\pm}/ \partial \eta = 0$ with $\eta \in [0,1]$.
Since the form of the probability profile is rigorous\cite{01},
we obtain the correct values of $\eta$, satisfying the equation
\begin{equation}
\frac{8}{\chi^{\pm}} = 
\frac{\eta (1 - \eta^2)(3\eta^2+5)}{{\rm sgn}(E) - 
( \frac{\epsilon}{2} \pm 1) \eta} = f_{\pm}(\eta).
\end{equation}
From Eq.~(10) it is clear that the states will appear above the band for
$\chi^{\pm} > 0$ and below the band for $\chi^{\pm} < 0$.
[Note that the band edges are at $E/V = \pm 2$.]
We will consider positive values of $\chi^{\pm}$ and, hence, the
sgn($E$) may be taken to be positive.
For symmetric states,
the value of $f_{+} ( \eta )$ exhibits the extreme values:
$f_{+} ( \eta ) \rightarrow 0$ as $\eta \rightarrow 0$ and
$f_{+} ( \eta ) \rightarrow \infty$ as
$\eta \rightarrow 1 / ( \epsilon /2 + 1 )$.
Thus, $f_{+} ( \eta )$ increases as $\eta$ increases from $\eta = 0$
and diverges at $\eta = 1 / ( \epsilon /2 + 1)$,
and it becomes negative for $\eta > 1 / ( \epsilon /2 + 1)$.
This means that there will always be only one solution (symmetric state)
for all positive values of $\epsilon$ and $\chi$.
For antisymmetric states, we found that
$f_{-} ( \eta ) \rightarrow 0$ as $\eta \rightarrow 0$ and
$f_{-} ( \eta ) \rightarrow \infty$ as
$\eta \rightarrow 1 / ( \epsilon / 2 - 1)$.
The divergence of $f_{-} ( \eta )$ is, however, confined between 0 and 1
if and only if $\epsilon > 4$.
Therefore, for $\epsilon > 4$, $f_{-} ( \eta )$ increases
as $\eta$ increases from $\eta = 0$,
diverges at $\eta = 1 / ( \epsilon /2 - 1)$, and finally becomes negative
for $\eta > 1/ ( \epsilon /2 - 1)$.
This implies that there will always be one antisymmetric state
for $\epsilon > 4$ for any value of $\chi > 0$.
However, the situation is different for $\epsilon < 4$;
the value of $f_{-} ( \eta )$ approaches 0 as $\eta \rightarrow 0$
and $\eta \rightarrow 1$ and remains positive and finite for $0< \eta <1$.
This implies that $f_{-} ( \eta )$ has at least one maximum in the range of
$0 < \eta <1$.
We confirm by a graphical analysis that there is only one maximum.
Therefore, there will be a critical value of $\chi$ separating
two antisymmetric states from the region of no antisymmetric state
for $\epsilon < 4$.
The dashed line in Fig.~1 is such a critical line and is denoted by
$\chi_{cr}^{a}$.
Thus, there are three regions in the entire plane of $\epsilon$ versus
$\chi$, having different number of symmetric and antisymmetric SL states.
The region of $\epsilon > 4$ has one symmetric and one antisymmetric SL
states, while the region of $\epsilon < 4$ is divided into two regions
by the critical line $\chi_{cr}^{a}$,
below of which has only one symmetric state and above of which has
one symmetric state and two antisymmetric states.
Therefore, the maximum number of SL states contributed by the symmetric and
anti-symmetric solutions is three.

Next, we consider the asymmetric solutions, i.e., for $\beta \ne \pm 1$.
The effective Hamiltonian in Eq.~(8) should be considered as a function of
two variables, $\eta$ and $\beta$,
while $\epsilon$ and $\chi$ remain constant for a particular system.
The fixed-point solutions corresponding to the SL states
can be obtained from the following two extreme conditions
\begin{eqnarray}
\partial H_{\rm eff} / \partial \eta & = & 0, \nonumber \\
\partial H_{\rm eff} / \partial \beta & = & 0 .
\end{eqnarray}
After a trite algebra, the fixed-point equations reduce to
\begin{eqnarray}
2 ( 1 + \beta^2 ) ( 1 + \beta^{2} - 2 \eta \beta  - \epsilon \eta ) \nonumber \\
- \chi \eta (1- \eta^{2})(3\beta^4\eta^2+3\beta^4+2\beta^2)  =  0 
\nonumber \\
\noindent {\rm and}~~~~~~~~~~~~~~~~~~~~~~~~~~~~~~~~~~~~~~~~~~~~~~~~~~ \nonumber \\
2(1+\beta^2)(1-\beta^2-\epsilon\beta) \nonumber \\
+ \chi\beta(1-\eta^2)
(2\beta^2\eta^2+3\beta^2+1)=0
  \end{eqnarray}
These coupled equations can not be solved analytically and, accordingly,
we analyze them numerically to find the number of asymmetric SL states
(number of solutions with $\eta \in [0,1]$).

\begin{center}
\epsfig{file=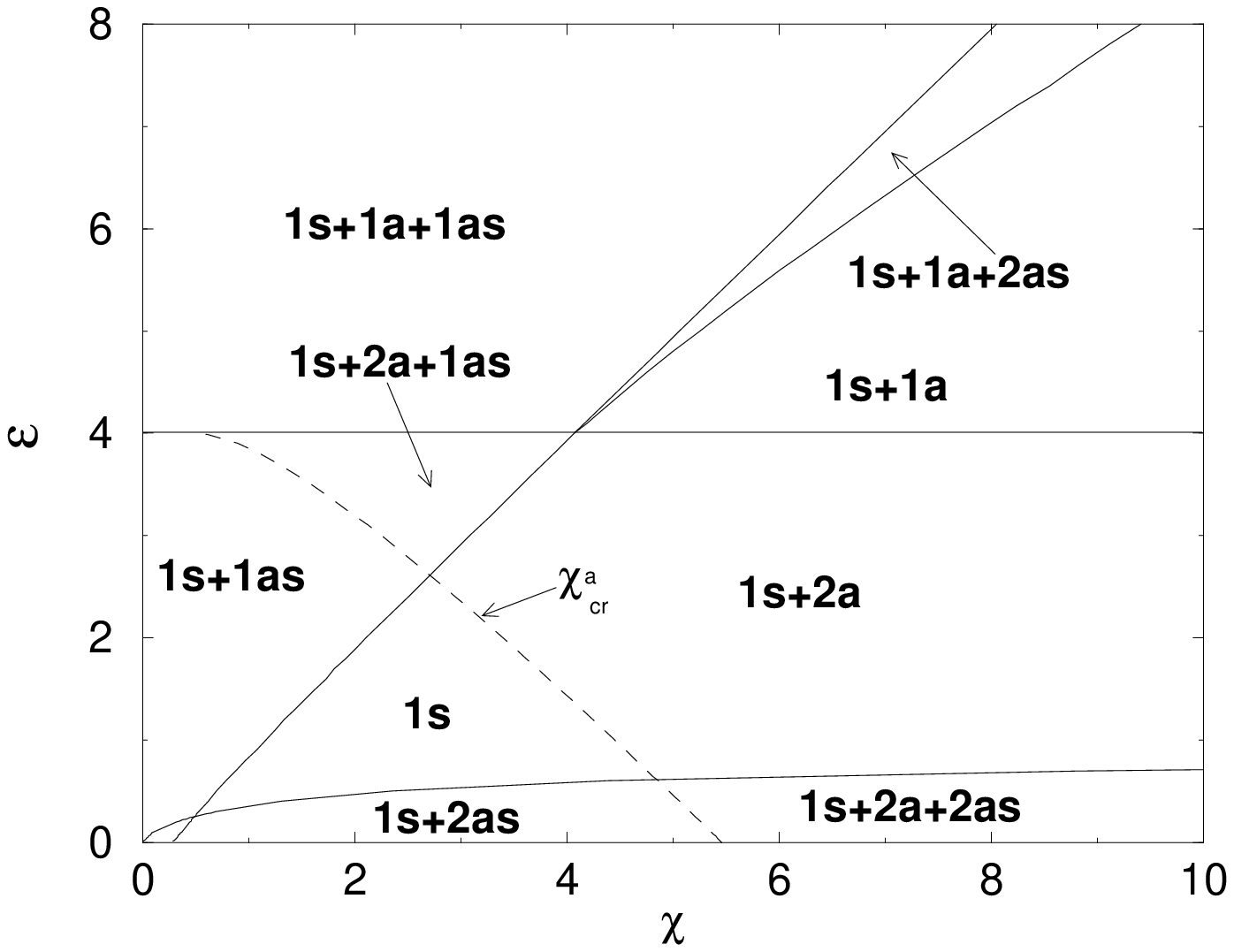,width=7.8cm,height=6.2cm}
\begin{figure}[bh]
\caption{The phase diagram for SL states in the plane of
$\epsilon$ versus $\chi$ for a system of a one-dimensional chain
with a linear impurity and a modified nonlinear impurity.
Different regions are marked with the number of states
and the kind of states; the region marked by ``1s+2a+2as'' means
that the region has one symmetric, two antisymmetric and two 
asymmetric SL states. The small unmarked region has only one 
symmetric SL state.}
\label{fig1}
\end{figure}
\end{center}
\vspace{0.2cm}

It is found that there are three critical lines separating
different number of asymmetric SL states in the plane of 
$\epsilon$ versus $\chi$ denoted by the solid curves in Fig.~1.
Maximum number of asymmetric SL states obtained is two.
Considering all possible states we find that maximum number
of SL states is five.
The ``s'', ``a'' and ``as'' (in Fig.~ 1) stand for symmetric, 
antisymmetric and asymmetric states, respectively.
For example, the region marked as ``1s+2a+2as'' has one symmetric SL state,
two antisymmetric SL states, and two asymmetric SL states.
It should be noted that the number of states changes one by one as
we go from one region to another in the phase space.
As an example, as we cross from the region ``1s+1as'' to the region
``1s+2a+1as'', the number of states increases by two;
however, the number of states on the critical line which separates
two regions is ``1s+1a+1as'', yielding the increment of the number of
states one by one.
The maximum number of SL states in one dimensional chain with two
linear impurities was found to be two \cite{01} and that in one 
dimensional chain with two modified nonlinear impurities
was found to be four\cite{22}.
We thus conclude that one linear impurity and one modified nonlinear 
impurity play a role to increase the maximum number of SL states.
The phase diagrams of SL states here is very complicated when
compared to the that for one dimensional chain with two modified nonlinear
impurities.

Figure~2 shows the variation of energy of the states
as a function of $\chi$ for a fixed value of $\epsilon = 6$.
We see that three SL states appear at $\chi=5$ as expected from
the phase diagram (Fig.~1). The energy of 
the symmetric state (top solid line in Fig. ~2) and that of the
antisymmetric state (another solid line in Fig. ~2) increases
while the energy of the asymmetric state decreases as $\chi$ 
increases. At $\chi \approx 6.06$ one more asymmetric state appear 
whose energy increases with the increment of $\chi$ and finally
both the asymmetric states disappear at $\chi = 6.55$. Thus we
find that the symmetric state and the anti symmetric state are
strongly localized compared to the asymmetric states in the 
parameter regime considered in Fig.~2. Through stability analysis
\cite{22} it can be shown that the state whose energy decreases
with the increment of $\chi$ is unstable.

\vspace{0.4cm}
\begin{center}
\epsfig{file=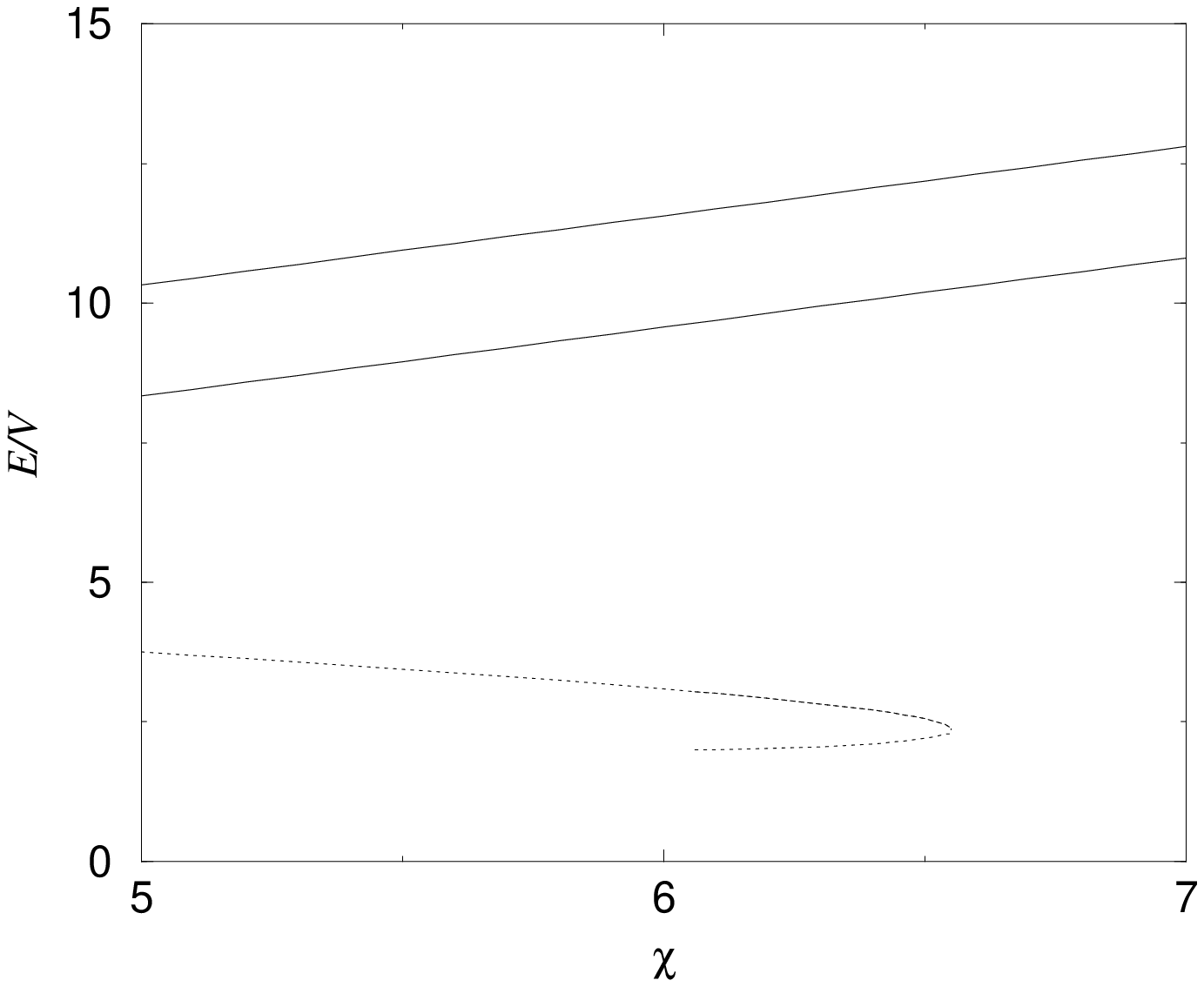,width=7.4cm,height=5.7cm}
\begin{figure}[bh]
\caption{Variation of energy as a function of $\chi$ is shown for
         $\epsilon = 6$.}
\label{fig2}
\end{figure}
\end{center}

\section{Linear and modified nonlinear site energies at alternate sites}
In this section, we consider a one-dimensional chain with alternate
site energies; the even sites have a linear energy $\epsilon$ and
the odd sites have a modified nonlinear energy 
$\chi (|C_{n+1}|^{2} + |C_{n-1}|^2 + 2|C_n|^2)$.
Such a system can be described by the Hamiltonian given as
\begin{eqnarray}
H  =  \frac{1}{2} \sum_{n= - \infty}^{\infty} 
( C_{n} C_{n+1}^{\star} + C_{n}^{\star} C_{n+1} ) 
 + \frac{\epsilon}{2} \sum_{n= - \infty}^{\infty} | C_{2n}|^{2} \nonumber \\
+ \frac{\chi}{4} \sum_{n= - \infty}^{\infty} (|C_{2n+2}|^2 + |C_{2n}|^2
+ 2|C_{2n+1}|^2) |C_{2n+1}|^2.
\end{eqnarray}
For $\chi = 0$, the system reduces to a periodically modulated linear system.
The translational symmetry of the system is preserved for both
$\chi = 0$ and $\chi \ne 0$.
Using $C_{n} = \phi_{n} \exp (- iEt)$,
we obtain the Hamiltonian in terms of $\phi_{n}$.
It is, however, not possible to find the exact ansatz for the localized
states in this system, although a certain rational ansatz can be considered.
For example, the on-site peaked solution, inter-site peaked or inter-site
dipped solutions are possible.
We will consider each of these in the subsequent subsections.

\vspace{0.4cm}
\begin{center}
\epsfig{file=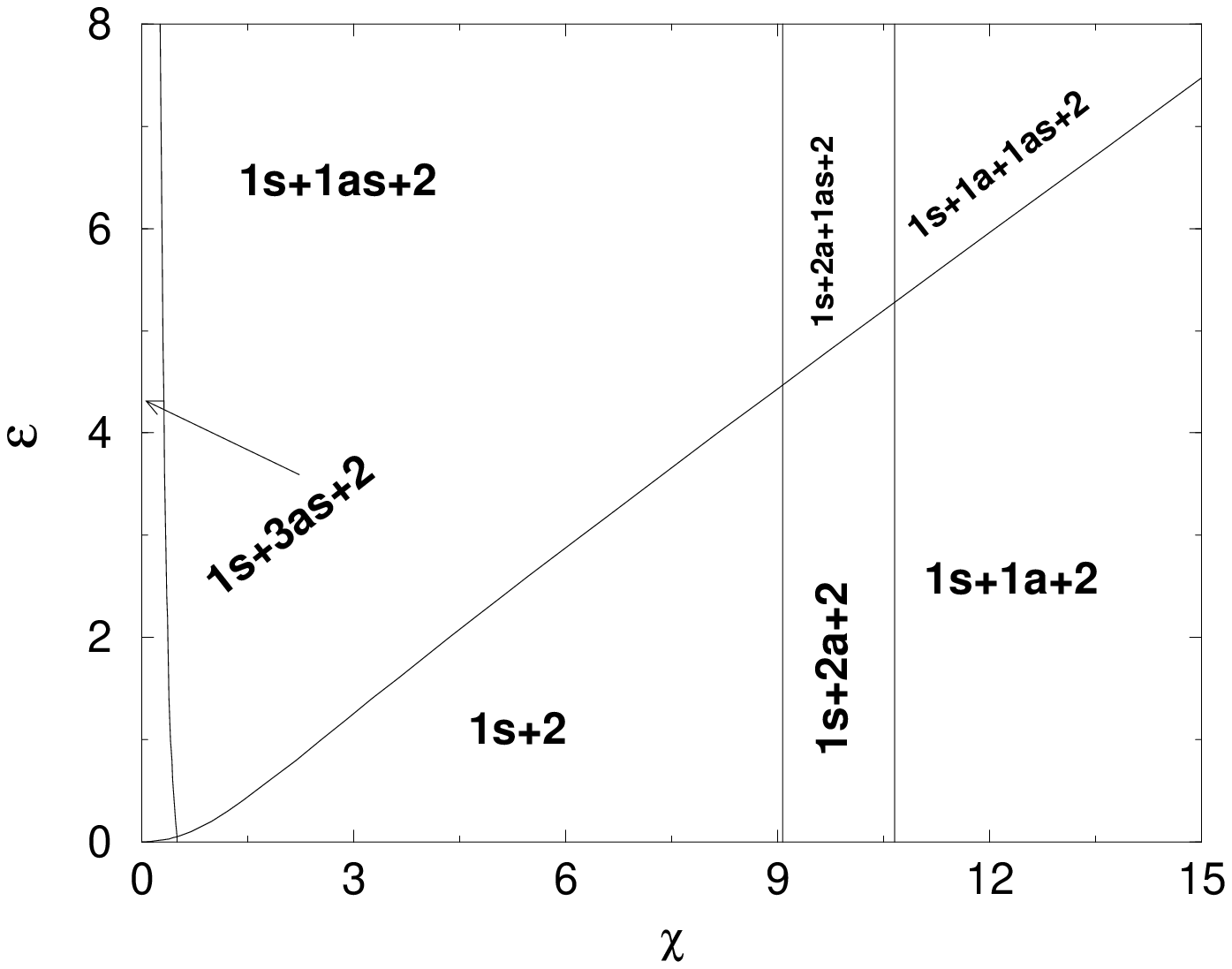,width=7.4cm,height=5.7cm}
\begin{figure}[bh]
\caption{ 
The phase diagram of SL states in the $\epsilon$ versus 
$\chi$ plane for the system of a one dimensional chain
with linear and modified nonlinear site energies at alternate sites.
The number `2' added in every region indicates that there are two
on-site peaked  states apart from the symmetric, antisymmetric and 
asymmetric states; one of them is peaked at the linear site and 
another one is peaked at the nonlinear site. For example, the region
marked by `1s+2a+1as+2' has one symmetric state, two antisymmetric 
states, one asymmetric state and two on-site peaked states (one 
peaked at the linear site and another one peaked at the nonlinear 
site).
}
\label{fig2}
\end{figure}
\end{center}

\subsection{On-site peaked solution}
Since the system has linear and nonlinear sites at regular intervals,
the solution may have a peak either at the linear site or at the nonlinear
site.
We first consider the solution peaked at the linear site, i.e. the
solution is peaked at the central site of the system.
We therefore consider the ansatz $\phi_{n} = \phi_{0} \eta^{|n|}$.
Using this and the normalization condition,
we obtain the effective Hamiltonian as
\begin{equation}
H_{\rm eff} = \frac{2 \eta}{1 + \eta^{2}} + \frac{\epsilon}{2} 
\frac{ (1 + \eta^{4}) }{ (1 + \eta^{2} )^{2}} + \frac{\chi}{2}
\frac{\eta^{2} (1 - \eta^{2} )}{ (1 + \eta^{2} )(1 + \eta^4 )}.
\end{equation}
The fixed-point equation, $\partial H_{\rm eff} / \partial\eta = 0$,
reduces to
\begin{eqnarray}
2(1-\eta^8)(1+\eta^4) - 2 \epsilon \eta (1-\eta^2)(1+\eta^4)^2  \nonumber \\
+ \chi\eta ((1+\eta^2)^2 (1-2\eta^2) (1+\eta^4) \nonumber \\ 
- \eta^2 (1-\eta^4) (1+2 \eta^2 + 3 \eta^4)) = 0.
\end{eqnarray}
Solving Eq.~(15), we found that there is always one SL state peaked
at the linear site for
any value of $\epsilon$ and $\chi$.
The localized solution peaked at the nonlinear site may be obtained
considering that each site is shifted by one lattice site along
either positive or negative direction, and the resulting
fixed-point equation is given as
\begin{eqnarray}
&2& (1 + \eta^2 ) ( 1 - \eta^4) ( 1 - \eta^8)^2 + 2 \epsilon \eta 
(1 - \eta^4) (1 - \eta^8)^2 \nonumber \\
&+& \chi \eta \left((1 + \eta^2) (1 - \eta^8)^2 (1 - 3 \eta^2) 
+8\eta^6 (1 - \eta^4)^2\right) \nonumber \\
&-&\chi\eta\left(4(1 - \eta^4)(1 - \eta^{16}) + 3 \eta^2 (1 - \eta^2) 
(1 - \eta^8)^2\right) = 0. 
\end{eqnarray}
It is found from the analysis of Eq.~(16) that only one SL state peaked 
at the nonlinear site appears irrespective of the values of $\chi$ and
$\epsilon$.

\subsection{inter-site peaked and dipped solutions}
For the inter-site peaked and dipped solutions we use the dimeric ansatz
in Eq.~(5).
After short calculations, we obtain the effective Hamiltonian for the
reduced dynamical system as
\begin{eqnarray}
H_{eff}  =  sgn(E) \eta 
+ \frac{\beta (1- \eta^{2} )}{1+ \beta^{2} } +  \frac{\epsilon}{2} \frac
{ (1+ \beta^{2} \eta^{2} ) }{ (1 + \beta^{2} ) ( 1+ \eta^{2} )} \nonumber \\ 
+ \frac{\chi}{4} \frac{(1-\eta^2)^2}{(1+\beta^2)^2}
\left[\frac{\eta^2(1+\beta^4)}{(1-\eta^8)} + \beta^2\right] \nonumber \\
+\frac{\chi}{4}\frac{(1-\eta^2)^2}{(1+\beta^2)^2}
\left[\frac{\eta^6(1+\beta^4)}{(1-\eta^8)} + \frac{(\beta^4+\eta^4)}
{(1-\eta^8)} \right]
\end{eqnarray}
with $\beta$ and $\eta$ as defined earlier.
We first consider the symmetric and antisymmetric states, i.e., $\beta = \pm1$.
The fixed-point equation yields
\begin{equation}
\frac{4}{\chi^{\pm}} = \frac{\eta\left((1+\eta^2)(1+2\eta^2-\eta^4) - 
\eta^4\right)}{(1 \mp \eta)( 1 + \eta^{2} )^{2} } 
= f^{\pm} ( \eta ) ,
\end{equation}
where ``+'' refers to the symmetric states and ``-'' to the
antisymmetric states.
It should be noted that $\epsilon$ has no role in the formation of
symmetric and antisymmetric SL states in the system.
From Eq.~(18) it is clear that only one symmetric state appears for any
finite value of $\chi$.
However, for antisymmetric states, the situation is different.
The value of $f^{-} ( \eta )$ in Eq.~(18) initially increases
as $\eta$ increases and finally decreases as $\eta$ approaches to 1,
leaving a maximum at $\eta \in [0,1]$.
Thus, there are two critical values of $\chi$ such that there
is no anti-symmetric SL state for $\chi < 9.07$,
two anti-symmetric SL states for $9.07 < \chi < 10.66$
and one anti-symmetric SL state for $\chi \ge 10.66$,
shown as the two vertical lines in Fig.~3.

We next consider the asymmetric states of $\beta \ne 1$.
In this case $\epsilon$ plays its role, unlike the case of symmetric and
antisymmetric states.
The fixed-point equations are given as
\begin{eqnarray}
(1-\eta^8)^2\left[(1+\eta^2)^2(1+\beta^2)(1+\beta^2-2\beta\eta) 
-\epsilon\eta(1-\beta^4)\right]  \nonumber \\
+ \chi\eta(1-\eta^4)^2(1-\eta^8) \nonumber \\
\left[\frac{1}{2}(1+\beta^4)(1+3\eta^4)
-\beta^2(1+\eta^2)(1+\eta^4) + \eta^2\right]  \nonumber \\
+ \chi\eta^3(1-\eta^4)(1-\eta^8)(1+\beta^4) \nonumber \\ 
\left[2\eta^6-(1+\eta^2)
(1+\eta^4)\right] \nonumber \\
- \chi\eta(1-\eta^4)^2(\beta^4+\eta^4)
(1+\eta^2+\eta^4-\eta^6) = 0 \nonumber \\
\end{eqnarray}
and
\begin{eqnarray}
(1-\eta^8)(1-\eta^4)(1-\beta^4) \nonumber \\
- \epsilon\beta (1-\eta^8)(1+\beta^2) (1-\eta^2) \nonumber \\
+ \chi\beta(1-\eta^2)(1-\eta^4)(1+\beta^2) \nonumber \\
\left[\frac{(1-\eta^8)}{2} + \beta^2(1+\eta^2+\eta^6)\right] \nonumber \\
-\chi\beta(1-\eta^2)(1-\eta^4)[\eta^2(1+\beta^4)(1+\eta^4) \nonumber \\
+ \beta^4(1-\eta^8) +\beta^4+\eta^4] = 0
\end{eqnarray}
Solving Eqs.~(19) and (20) we obtain two critical curves (solid curves
in Fig.~ 3) separating
the $\chi - \epsilon$ plane into various regions with various number
of asymmetric SL states as shown in Fig.~3. Symmetric states are
denoted by `s', antisymmetric states are denoted by `a' and asymmetric 
states are denoted by `as'. The region marked by
`1s+2a+1as+2' contains one symmetric SL state, two antisymmetric SL
states, one asymmetric SL state, one SL state peaked on the nonlinear
site and one SL state peaked at the linear site.
The maximum number of states obtained in this case is 6 which is the
same as for the chain with modified nonlinear impurities of different
strengths at alternate sites. Thus in this case the linear impurity
plays the same role as that of nonlinear impurity to obtain the maximum
number of SL states. No SL state is obtained for $\chi = 0$.
This is expected because the system with $\chi = 0$ reduces to a
periodically modulated linear system which again may be renormalized
to a perfect linear system.

\section{summary}

Formation of stationary localized states in one dimensional chain
due to the presence of a linear impurity and a modified nonlinear 
impurity is studied. Maximum number of SL states in this case is 
found to be five which is larger than that obtained for a one 
dimensional chain with two modified nonlinear impurities. Thus
one may conclude that the presence of both linear and nonlinear 
impurities plays a role to increase the maximum number of SL states.
The variation of energy of SL states as a function of the nonlinear
strength for a fixed value of linear impurity strength is presented.
It is found that the symmetric and antisymmetric states are 
strongly localized when compared with the asymmetric states. 

Furthermore, a one dimensional chain with linear and modified 
nonlinear energies at alternate sites is considered. Here the 
maximum number of SL states is found to be six. The linear energy
strength at alternate sites in the chain does not play any role
to produce the symmetric and antisymmetric SL states and
the system does not produce any SL state for $\chi$=0 which
is expected. However, surprisingly, the linear energy strength 
plays a role to produce the asymmetric SL states. Rich phase 
diagrams of SL states for both the systems considered here are
presented.
\acknowledgments
This work was supported by the Korea Research Foundation Grant
(KRF-2000-015-DP0101). The authors are grateful for the support.


\begin{thebibliography}{99}

\bibitem{01} E. N. Economou, {\it Green's Function in Quantum Physics}
             (Springer, Berlin, 1979).
\bibitem{02} M. I. Molina and G. P. Tsironis, Phys. Rev. B {\bf 47}
             15 330 (1993).
\bibitem{03} G. P. Tsironis, M. I. Molina, and D. Hennig, 
             Phys. Rev. E {\bf 50}, 2365 (1994).
\bibitem{04} M. I. Molina and G. P. Tsironis, Int. J. Mod. Phys. B 
             {\bf 9}, 1899 (1995).	 
\bibitem{05} Y. Y. Yui, K. M. Ng and P. M. Hui, Phys. Lett. A {\bf 200}, 
             325 (1995); Y. Y. Yui, K. M. Ng and P. M. Hui, 
	     Solid State Commun. {\bf 95}, 801 (1995). 	    
\bibitem{06} P. M. Hui, Y. F. Woo and W. Deng, Conden. Matt. {\bf 8},
             2011 (1996).
\bibitem{07} B. C. Gupta and K. Kundu, Phys. Rev. B {\bf 55}, 894 (1997).
\bibitem{08} B. C. Gupta and K. Kundu, Phys. Rev. B {\bf 55}, 11033 (1997).
\bibitem{09} K. Kundu and B. C. Gupta, Eur. Phys. J. B {\bf 3}, 23 (1998).
\bibitem{10} A. C. Scott and L. MacNeil, Phys. Lett. A {\bf 98} 87 (1983).
\bibitem{11} J. C. Eilbeck, P. S. Lomdahl and A. C. Scott, Physica D 
             {\bf 16}, 318 (1985)
\bibitem{12} R. S. MacKay and S. Aubry, Nonlinearity, {\bf 7} 1623 (1994).
\bibitem{13} A. B. Aceves, C. De Angelis, T. Peschel, R. Muschall,
             F. Lederer, S. Trillo and S. Wabnitz, Phys. Rev. E {\bf 53}, 
	     1172 (1996).
\bibitem{14} B. Malomed and M. I. Weinstein, Phys. Lett. A {\bf 220}, 
             91 (1996).	 
\bibitem{15} B. C. Gupta and K. Kundu, Phys. Lett. A {\bf 235} 176 (1997).
\bibitem{16} A. Ghosh, B. C. Gupta and K. Kundu, J. Phys. Conden. Matt. 
             {\bf 10}, 2701 (1998).
\bibitem{17} V. M. Kenkre, G. P. Tsironis and D. K. Campbell, {\it 
             Nonlinearity in Condensed Matter}, edited by A. R. Bishop
	     et al. (Springer-Verlag, Berlin, 1987), p. 226.
\bibitem{18} V. M. Kenkre and G. P. Tsironis, Phys. Rev. B {\bf 35}
             1473 (1987).
\bibitem{19} Bikash C. Gupta and Sang Bub Lee, (Unpublished).	     
\bibitem{20} G. Kopidakis, C. M. Soukoulis and E. N. Economou, Phys.
             Rev. B {\bf 49} 7036 (1994); {\bf 51} 15 038 (1995).
\bibitem{21} G. Kalosakas, G. P. Tsironis and E. N. Economou,
             J. Phys. Conden. Matt. {\bf 6} 7847 (1994).
\bibitem{22} Bikash C. Gupta and Sang Bub Lee, Phys. Lett. A 
             {\bf 269} 130 (2000).
\end{thebibliography}
\end{document}